\documentclass[a4paper,11pt,oneside]{article}
\usepackage{multicol}
\usepackage{epsfig}   
\usepackage{moreverb}  
\usepackage[none,bottom,dark,english]{draftcopy}
\draftcopyName{ TR 2004/162, CSSE, Monash U., .au }

\begin{document}

\title{Finding Approximate Palindromes in Strings Quickly and Simply.}
\author{Lloyd Allison \\
School of Computer Science and Software Engineering, \\
Monash University,
Clayton, Victoria, Australia 3800. \\
 \\
Technical Report 2004/162
 }

\date{23 Nov. 2004 (draft 19 Sept)}

\maketitle

\begin{multicols}{2}

\noindent


\noindent
Abstract:
Described are two algorithms to find long approximate palindromes in a string,
for example a DNA sequence.
A simple algorithm requires $O(n)$-space
and almost always runs in $O(k.n)$-time
where n is the length of the string and
k is the number of ``errors'' allowed in the palindrome.
Its worst-case time-complexity is $O(n^2)$ but this
does not occur with real biological sequences.
A more complex algorithm guarantees $O(k.n)$ worst-case time complexity.

Code of the simple algorithm will be placed at
http://www.csse.monash.edu.au/ $\sim$lloyd/tildeProgLang/Java2/Palindromes/



\section{Introduction}

An (exact) palindrome, $p$, is a string of symbols that reads the same
forwards and backwards, i.e.~either $p=w.w'$ or $p=w.c.w'$
where $w$ is a string, $c$ is a symbol and $w'=reverse(w)$;
for complementary palindromes in DNA (RNA)
we have $w'=reverse(complement(w))$
where A and T (U) are complementary, as are C and~G.
The first case, $p=w.w'$, is called an even-palindrome and
the second, $p=w.c.w'$, is called an odd-palindrome and
either the ``gap'' between $w$ and $w'$ or the symbol $c$
is called the {\it centre} of the palindrome.
Finding palindromes within a long string leads to
various classic computing problems,
e.g.~the {\it longest} palindrome within a string can be found in
linear-time by using a suffix-tree (Weiner 1973, McCreight 1976).

Palindromes can be interesting biologically (e.g.~Tsunoda 1999, Rozen 2003) and
reverse complementary palindromes are relevant to
hair-pin loops in RNA folding.
But, in biology, palindromes are often allowed to be {\it approximate}:
$k$~``errors'' or ``differences'' are allowed between $w$ and $reverse(w')$,
that is $w$ and $reverse(w')$ can have an edit-distance of~$k$.
Note that in general one approximate palindromic string, $p$, may correspond
to multiple decompositions $p=w.w'$ or $p=w.c.w'$
(it is not necessary that $|w|=|w'|$), and
a decomposition may correspond to multiple alignments of $w$ and $reverse(w')$;
we prefer a cheapest decomposition and alignment and require costs${\le}k$.

Porto and Barbosa (2002) gave an $(k^2 n)$-time algorithm to
find long approximate palindromes in a string.
This paper gives a simple algorithm to find long approximate palindromes.
It runs in $O(n)$-space and, almost always, in $O(k.n)$-time;
e.g.~for $k{\sim}10$, a million bases of real DNA can be processed
in a few seconds on a p.c., most of that time being for I/O.
A more complex algorithm guarantees $O(k.n)$ running time.

[[ fig 1 near here ]]

\section{Algorithm}

\begin{figure*}
\begin{center}
\epsfig{file=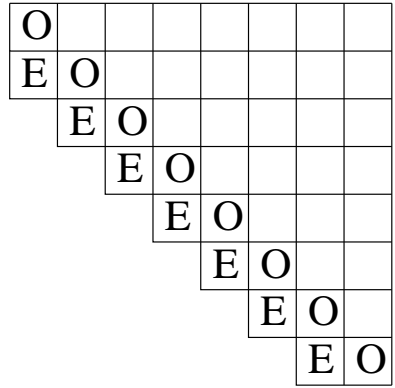}
\end{center}
\caption{Odd and Even Origins}
\label{EvenOdd}
\end{figure*}

\begin{figure*}
\begin{center}
\epsfig{file=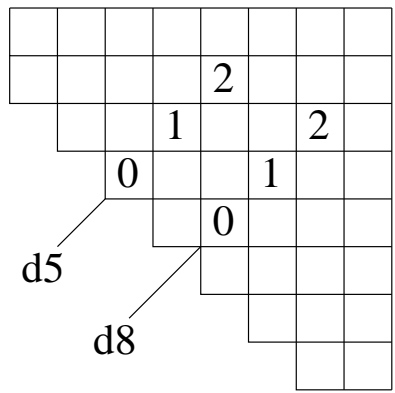}
\end{center}
\caption{Some Distances along Diagonals}
\label{Distances}
\end{figure*}

It is convenient to describe the simple algorithm in terms of
a {\it distance matrix}, related to those used in some alignment algorithms.
The matrix is a variation on a triangular matrix (Figure~\ref{EvenOdd}).
An odd exact palindrome is centered on one of the cells marked `O', and
an even exact palindrome on one of the cells marked~`E'.
A marked cell is called an {\it origin}.
Diagonals that run from an origin in a NE direction are important;
note that an odd exact palindrome corresponds to an even-numbered diagonal and
an even exact-palindrome to an odd-numbered diagonal.
Also important are distances along diagonals (Figure~\ref{Distances}).
It must be pointed out that the algorithm does not directly use a
distance matrix; rather it operates on a different but equivalent matrix
to be described.

[[ fig 2 near here ]]

\begin{figure*}
\begin{center}
\epsfig{file=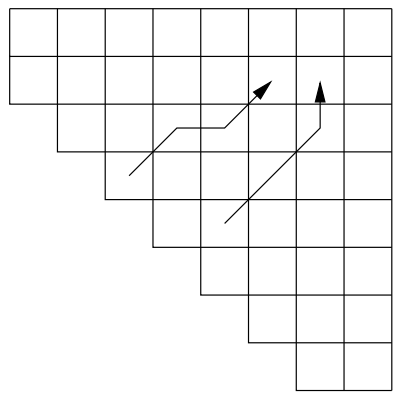}
\end{center}
\caption{Example Paths}
\label{Paths}
\end{figure*}

An {\it approximate} palindrome, $p$, {\it together} with an alignment of
$w$ and $reverse(w')$ where $p=w.w'$ or $p=w.c.w'$, implying a cost,
is equivalent to a path (Figure~\ref{Paths}) which
extends step by step N, E, and/or NE, some distance from an origin.
A NE step represents a match or a mismatch. N~and E steps represent indels.
Each cell of the (notional) distance matrix holds the minimum cost of
some optimal path from some origin, not necessarily on the same diagonal,
to the cell.
The position of any cell in the distance matrix specifies
an approximate palindrome, $p$ itself, {\it without} any associated alignment;
the position fixes the start and the end of the string~$p$.
Obviously we want the minimum-cost for an approximate palindrome.

[[ fig 3 near here ]]

The algorithm actually uses a different but equivalent matrix, $reach[d][e]$,
indexed by $d$ which corresponds to a diagonal-number in the distance matrix,
and by ``error'' count, $e$, where $0 \le e \le k$.
$reach[d][e]$ holds the maximum distance along diagonal $d$ of
the distance matrix that can be reached by an approximate palindrome for
a cost of at most~$e$.

The algorithm initially finds exact palindromes, $e=0$,
i.e.~paths that move NE only, as long as this can be done for a cost of zero.
It then iterates over the number of errors allowed, $e=1..k$, and,
within that, over diagonal-number,~$d$, where it executes the general step:
\begin{verbatim}

reach[d][e] =
 max(reach[d-1][e-1]+x,
     reach[d  ][e-1]+1,
     reach[d+1][e-1]+x), where x=d & 1;

while endsMatch(d,reach[d][e]) do
  reach[d][e]++  // extend for free
\end{verbatim}
It is an instance of a {\it greedy} strategy (e.g. Ukkonen 1983).
Other tests, not shown, check that the ends of the string are not overrun.
On termination, $reach[d][k]$ holds the maximum NE-erly distance from an origin
of an acceptable path {\it ending} on diagonal~$d$, thus
giving long approximate palindromes.

$O(n)$-space is sufficient to find the approximate palindromes because
$reach[~][e]$ only depends on $reach[~][e-1]$.
If alignments (paths) are also required, either $O(k.n)$-space
is required to keep all of $reach[~][~]$ or,
probably more sensibly assuming path lengths~$<<n$,
paths can be recovered later by a separate process.

The simple algorithm's worst-case behaviour, $O(n^2)$-time,
is for strings such as $A^n$, $(AT)^{n/2}$, and similar.
The cause is looping in order to check a run of matches to
extend a path directly NE for zero cost;
in practice the {\it average} run ends quickly on real DNA sequences.
The complex algorithm is, in principle, formed by
replacing the simple algorithm's loop by a constant-time step
(following linear-time preprocessing) which uses a suffix-tree and
a least-common-ancestor (LCA) algorithm such as
that of Bender and Farach-Colton (2000).

\section{Results}

The simple algorithm was coded in Java and tested on a Linux p.c.,
AMD Athlon XP{\texttrademark} 2400+ processor, 512MB of memory.
It confirmed $O(k.n)$-time complexity in practice on real DNA,
e.g.~processing chromosome~3 (1.06Mb) of the malaria organism
{\it Plasmodium falciparum} (Gardner et al 2002)
as follows: $k=10$ in $8.0s$, $k=20$ in $10.2s$, $k=40$ in~$14.3s$.
Such DNA is approximately $80\%$ AT-rich and is the kind of {\it real} DNA
most likely to cause problems for this kind of algorithm if any will.
The algorithm has not been observed to make more than $3.7(k+1)n$
symbol comparisons on real DNA sequences.

\section{References}

\begin{description}
\itemsep=0in
\parsep=0in

\item
Bender, M. A. and Farach-Colton, M. (2000)
The LCA problem revisited.
Proc. of the 4th Latin American Symp. on Theoretical Informatics, pp.88--94.

\item
Gardner M. J., et al (2002)
Genome sequence of the human malaria parasite Plasmodium falciparum.
Nature 419 pp.498--511.

\item
McCreight, E. M. (1976)
A space-economic suffix tree construction algorithm.
J. of the ACM 23(2) pp.262--272.

\item Porto, A. H. L. and Barbosa V. C. (2002)
Finding approximate palindromes in strings.
Pattern Recognition 35 pp.2581--2591.

\item
Rozen S., Skaletsky, H. et al (2003)
Abundant gene conversion between arms of palindromes in human and
ape Y-chromosomes.
Nature 423 pp.873--876.

\item
Tsunoda T., Fukagawa M. and Takagi T. (1999)
Time and memory efficient algorithm for extracting palindromic and
repetitive subsequences in nucleic acid sequences.
Pacific Symp. on Biocomputing 4 pp.202--213.

\item
Ukkonen, E. (1983)
On approximate string matching.
Proc. Int. Conf. on Foundations of Computation Theory,
Borgholm, Sweden, pp.487--495.

\item
Weiner, P. (1973)
Linear pattern matching algorithms.
14th IEEE Symposium on switching and automata theory pp.1--11.

\end{description}

\end{multicols}

\newpage

\end{document}